# Higher-Order-Phonon Scattering Governs Targeted Control of Heat Conduction in Bulk Boron Arsenide

Tianhao Li[1], Yangjun Qin[1,2], Dongkai Pan[1], Shixian Liu[3], Han Meng[2*], Nuo Yang[2*]

[1] School of Energy and Power Engineering, Huazhong University of Science and Technology, Wuhan 430074, China

[2] School of Science, National University of Defense Technology, Changsha 410073, China

[3] Independent Researcher, Xinzhou 034100, China

[*] Corresponding author: menghan@nudt.edu.cn (H. M.), nuo@nudt.edu.cn (N. Y.)

## Abstract

Conventional thermal conductivity modulation relies on structural modifications, precluding reversible in situ regulation. Targeted phonon excitation, a promising strategy for dynamic heat conduction control via selective perturbation of phonon populations and scattering channels, has only been demonstrated in two-dimensional systems. Here, using boron arsenide (BAs) as a model system, we show that higher-order phonon scattering governs targeted thermal control in three-dimensional bulk materials. With a large acoustic–optical phonon gap, BAs has limited three-phonon scattering phase space, rendering four-phonon scattering a significant higher-order anharmonic channel. Using first-principles calculations and phonon Boltzmann transport analysis, we find that targeted excitation modulates BAs thermal conductivity in a strongly frequency-dependent manner. Within the three-phonon-only framework, the modulation at 300 K is weak but clearly bidirectional. However, including four-phonon scattering qualitatively changes the modulation to predominantly suppressive. In the combined 3ph+4ph framework, the strongest suppression occurs at 20.5 THz, with $\kappa/\kappa 0$ dropping to 0.828 and 0.415 for excitation intensities of 5 and 25, respectively. Scattering-rate analysis reveals that four-phonon scattering elevates the intrinsic scattering background and enhances excitation-induced scattering of low-frequency heat-carrying phonons, dictating the net response. Furthermore, comparing 300 K and 100 K shows that lowering temperature weakens four-phonon scattering dominance and partially restores 3ph-like bidirectional modulation. These findings establish higher-order phonon scattering as a governing factor for bulk targeted thermal control and guide phonon engineering and dynamic thermal management in 3D systems.

## Introduction

Thermal conductivity, which quantifies the ability of a material to conduct heat, plays a vital role in numerous technological applications[1-4]. Materials with high thermal conductivity, such as graphene and boron nitride, are ideal for heat dissipation in electronic devices[5], whereas low thermal conductivity is crucial for achieving high thermoelectric figure of merit (zT) and energy conversion efficiency[6-8]. The capability to precisely modulate thermal conductivity is therefore essential for addressing these contrasting demands.

Conventional strategies for tuning thermal conductivity[9-14] including nanostructuring[15-17], doping[18-20], introducing disorder[21-23], and applying strain[24,25] primarily rely on manipulating phonon transport through structural modifications. For instance, Wan et al. minimized the thermal conductivity of graphene nanoribbons through resonance-hybridized nanopillar designs that optimize phonon transport[26], while Chen et al. significantly enhanced polyethylene thermal conductivity via stretch-induced polymer chain alignment[27]. Although effective, these methods typically induce irreversible structural changes, preventing dynamic and reversible thermal modulation. Consequently, there is a pressing need for thermal conductivity control mechanisms that preserve intrinsic material structure.

Recently, targeted phonon excitation has emerged as a promising strategy for modulating thermal transport by selectively perturbing phonon populations, scattering, and mode coupling without introducing permanent structural changes[28-31]. In two-dimensional materials, this idea has already shown remarkable potential: graphene exhibits a 28% enhancement or 49% suppression in thermal conductivity[30], while hexagonal boron nitride (hBN) shows tunability ranging from 30% enhancement to 60% suppression[28]. In parallel, terahertz phonon engineering in van der Waals heterostructures has further highlighted the unusual ability of low-dimensional systems to generate, manipulate, and confine nonequilibrium phonon excitations[29]. Experimental studies have also begun to demonstrate the feasibility of selective phonon control; for example, direct terahertz excitation of low-energy phonons was shown to prolong hot-carrier cooling in halide perovskites[31], and intense terahertz pulses were found to excite a Raman-active A1g phonon in bismuth[32]. These advances collectively suggest that targeted phonon excitation provides a fundamentally new route for in situ control of heat transport beyond conventional structure-modification strategies. However, whether this strategy can be extended to three-dimensional bulk materials, where phonon coupling is more complex, remains unclear.

Boron arsenide (BAs), with an ultrahigh thermal conductivity[33-36] and weak anharmonicity, provides an ideal platform to explore targeted phonon excitation in 3D bulk systems. Unlike many conventional bulk crystals in which three-phonon scattering usually dominates thermal resistance and higher-order scattering is often a minor correction, BAs has a strongly restricted three-phonon scattering phase space associated with its large acoustic–optical phonon gap. As a result, four-phonon scattering contributes appreciably to phonon relaxation and can substantially modify its thermal conductivity. This unusual balance between weak three-phonon scattering and non-negligible four-phonon scattering makes BAs particularly suitable for examining how higher-order anharmonicity affects targeted-excitation modulation in bulk systems.

This work takes bulk boron arsenide (BAs) as a representative material to examine how higher-order phonon scattering governs targeted control of heat conduction in a three-dimensional bulk system. Based on first-principles calculations and phonon

Boltzmann transport analysis, we first map out the frequency-dependent modulation behavior of thermal conductivity in BAs at room temperature within both the three-phonon-only and the combined three-phonon plus four-phonon frameworks. By comparing these two frameworks, we clarify how higher-order four-phonon scattering modifies the net modulation response from weak bidirectional tuning to predominantly suppressive thermal-conductivity modulation. In addition, the temperature dependence of the 3ph+4ph modulation is examined by comparing the results at 300 K and 100 K. Finally, the microscopic origin of the modulation is analyzed from the perspective of intrinsic scattering background and excitation-induced scattering enhancement of low-frequency heat-carrying phonons.

## Methods

First-principles calculations were performed using density functional theory (DFT) with the projector augmented wave (PAW) method[37,38] as implemented in the Vienna Ab initio Simulation Package (VASP). The PAW PBE pseudopotential was employed for all elements with a plane-wave cutoff energy of 400 eV. The exchange-correlation functional was treated within the Local Density Approximation (LDA), with an energy convergence threshold of $1\times10^{-8}$ eV. A $10\times10\times10$ $\Gamma$-centered k-point mesh was used for Brillouin zone sampling in the primitive cell. The lattice constants and internal atomic positions were fully relaxed until atomic forces were below $1\times10^{-9}$ eV/Å. The optimized lattice constant of BAs is 4.742 Å, consistent with the experimental value of 4.777 Å[35,39].

For our in-house three-phonon calculations, the harmonic and third-order interatomic force constants (IFCs) were obtained from first principles using the Phonopy package[40]. The harmonic IFCs were computed by employing density functional perturbation theory (DFPT) and the finite displacement method with a $5\times5\times5$ supercell, in which all possible interactions were included. Third-order IFCs were obtained using the finite displacement method with a $4\times4\times4$ supercell, considering interactions up to the fifth nearest neighbor. The three-phonon lattice thermal conductivity was calculated by iteratively solving the linearized phonon Boltzmann transport equation (BTE) using ShengBTE[41,42]. A Γ-centered 30×30×30 q-mesh was used to ensure sufficient convergence.

Because four-phonon scattering was expected to play an important role in BAs[43,44], the subsequent targeted-excitation calculations were extended from a three-phonon framework to a combined three-phonon plus four-phonon framework. These calculations were carried out using the open-source FourPhonon package[42], which was implemented as an extension of ShengBTE[41]. For the calculations including four-phonon scattering, we adopted the publicly available BAs crystal structure together with the accompanying force-constant files provided in the package example, including the fourth-order IFCs. In this framework, the three-phonon contribution was treated through the iterative solution of the linearized phonon Boltzmann transport equation (BTE), whereas the four-phonon scattering rates were included at the single-mode relaxation time approximation (RTA) level.

The lattice thermal conductivity was evaluated within the phonon BTE framework using ShengBTE package[41,42]. The expression is given by

$$\kappa_l^{\alpha\beta} = \frac{1}{k_B T^2 \Omega N} \sum_{\lambda} f_0(f_0+1)(\hbar\omega_\lambda)^2 v_\lambda^\alpha v_\lambda^\beta \tau_\lambda$$

where $k_B$ is the Boltzmann constant, $\Omega$ is the volume of the unit cell, $T$ is the absolute temperature, $N$ is the number of q-points sampling the Brillouin zone, $f_0$ is the Bose–Einstein distribution function $f_0(\omega_\lambda) = \frac{1}{\exp(\hbar\omega/k_B T)-1}$, $\hbar$ is the reduced Planck constant, $\omega_\lambda$ is the angular frequency of phonon mode $\lambda$, $v_\lambda^\alpha$ and $v_\lambda^\beta$ are the group velocity components of phonon mode $\lambda$ along the $\alpha$ and $\beta$ directions, respectively, and $\tau_\lambda$ is the lifetime of phonon mode $\lambda$. For calculations including four-phonon scattering, the total scattering rate was determined according to Matthiessen's rule as

$$\tau_\lambda^{-1} = \tau_{\lambda,3ph}^{-1} + \tau_{\lambda,4ph}^{-1} + \tau_{\lambda,iso}^{-1}$$

where $\tau_{\lambda,3ph}^{-1}$, $\tau_{\lambda,4ph}^{-1}$ and $\tau_{\lambda,iso}^{-1}$ denote the scattering rates arising from three-phonon, four-phonon, and isotope scattering, respectively. The four-phonon contribution was further written as

$$\tau_{\lambda,4ph}^{-1} = \tau_{\lambda,(++)}^{-1} + \tau_{\lambda,(+-)}^{-1} + \tau_{\lambda,(--)}^{-1}$$

where (++), (+−), and (−−) represent the recombination, redistribution, and splitting channels of four-phonon scattering, respectively.

Targeted phonon excitation was modeled as a continuous increase in the population of selected phonon modes. This was implemented by introducing a multiplier N into the phonon distribution function. For phonon modes within the targeted frequency window, the distribution was written as

$$f_\lambda = Nf_0 = \frac{N}{\exp(\hbar\omega/k_B T) - 1}$$

whereas all other phonon modes retained the intrinsic Bose-Einstein distribution. The thermal conductivity under targeted excitation was then recalculated within the same phonon-transport framework using the modified nonequilibrium phonon distribution. For the calculations including four-phonon scattering, both the three-phonon and four-phonon contributions to the total scattering rate were taken into account.

Because direct 3ph+4ph calculations for every targeted-excitation case were computationally prohibitive, we further employed the sampling and maximum-likelihood-estimation (sampling-MLE)[45] approach to accelerate the evaluation of phonon scattering rates and the resulting thermal conductivity. In this approach, a subset of three-phonon and four-phonon scattering processes was sampled for each phonon mode, and the corresponding modal scattering rates were directly estimated under RTA instead of exhaustively enumerating all allowed scattering processes. A Γ-centered 16×16×16 q-mesh was used for the calculations including four-phonon scattering.

## Results and Discussion

The harmonic phonon properties of bulk BAs, including the phonon dispersion and density of states (DOS), were first examined to identify suitable frequency ranges for targeted phonon excitation. As shown in Fig. 1a, our calculated phonon dispersion agreed well with previously reported theoretical and experimental results, confirming the reliability of the calculations[34,44]. The phonon dispersion exhibited an obvious band gap spanning approximately 10 to 19 THz. Correspondingly, the phonon DOS showed a dominant distribution over low-frequency acoustic and high-frequency optical modes. This band gap not only reduced the phonon group velocity by compressing the phonon bands, but also significantly suppressed the coupling between acoustic and optical branches by reducing the phonon scattering phase space. These results suggested that 0–10 THz and 19–22 THz were candidate frequency domains for targeted phonon excitation.

To identify the phonon modes that dominated thermal transport, which will be excited to modulate thermal conductivity, the phonon scattering rates and the spectral thermal conductivity $\kappa(\omega)$ were further calculated at 300 K. As shown in Fig. 1b, the scattering rates of low-frequency acoustic phonons were significantly lower than those of high-frequency optical phonons, indicating weaker scattering and thus longer phonon lifetimes of acoustic phonons. Consequently, the spectral thermal conductivity shown in Fig. 1c was dominated by low-frequency acoustic phonons, with a pronounced peak in the range of 4–8 THz, whereas the contribution of optical phonons was negligible. These results suggested that exciting low-frequency acoustic phonons was more promising and efficient for modulating thermal transport in bulk BAs than exciting high-frequency optical phonons. The large acoustic–optical phonon gap also gives BAs an unusual scattering hierarchy compared with many conventional bulk crystals. The restricted three-phonon scattering phase space leaves higher-order four-phonon processes with a non-negligible contribution to phonon relaxation, making BAs a suitable system for examining the role of higher-order phonon scattering in targeted control of heat conduction.

In addition, the intrinsic thermal conductivity of bulk BAs at 300 K is calculated to be 2302 W/(m • K) in the 3ph-only framework, while it decreases to 1220 W/(m • K) when four-phonon scattering is further included. This significant reduction directly indicates the important role of four-phonon scattering in phonon transport in BAs and further justifies the separate analyses based on the 3ph-only and 3ph+4ph frameworks in the following sections.

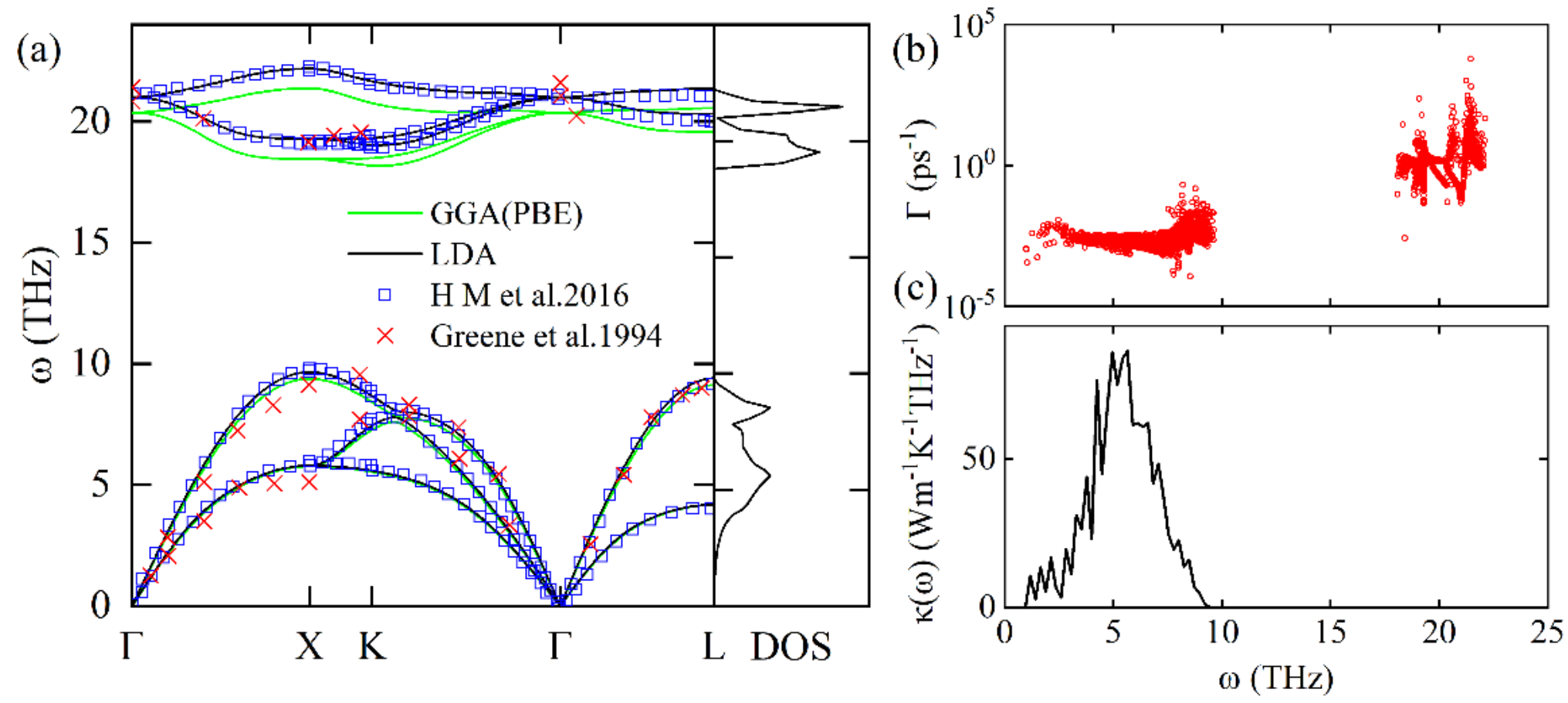

**Figure 1** (a) Phonon dispersion relations and density of states of bulk BAs at ground state. (b) Phonon scattering rates. (c) Spectral thermal conductivity of BAs at 300K.

After obtaining the dominant phonon modes, they are successively excited for thermal conductivity modulation. Firstly, the thermal conductivity of bulk BAs at 300 K was examined under targeted excitation within solely the three-phonon framework across the accessible phonon frequency range. In each calculation, only phonons within a 0.1 THz-wide frequency window were excited, while all other phonon modes retained their intrinsic distribution. Two excitation intensities, namely 5 and 25, were considered, and the modulation effect was quantified by the relative thermal conductivity, $\kappa/\kappa_0$.

As shown in Fig. 2a, targeted phonon excitation in the 3ph-only framework produced a weak but clearly bidirectional modulation of thermal conductivity. Over most of the scanned spectrum, the modulation remained limited in magnitude, but distinct enhancement and suppression windows could still be identified. In particular, the response near the upper edge of the acoustic branches, approximately in the 6–9 THz range, was especially sensitive to excitation intensity. At an excitation intensity of 5, a slight enhancement appeared around 8.3–8.9 THz, where $\kappa/\kappa_0$ slightly exceeded 1. When the excitation intensity was increased to 25, this enhancement was no longer retained in the same frequency range, and the modulation there became suppressive. Meanwhile, a weak enhancement survived only in a narrower interval of about 6.4–7.0 THz, where $\kappa/\kappa_0$ reached approximately 1.02.

These results showed that, within the three-phonon-only framework, targeted phonon excitation not only lead to a suppressive response, but also yielded a weak bidirectional modulation whose sign and magnitude depended sensitively on both the targeted frequency and the excitation intensity.

Since the intrinsic thermal conductivity is substantially reduced from 2302 W/(m • K) in the 3ph-only framework to 1220 W/(m • K) in the 3ph+4ph framework, four-phonon scattering is expected to play an important role in the targeted-excitation response of bulk BAs[36,43]. Therefore, the same frequency-dependent excitation calculations were further performed within the combined three-phonon plus four-phonon framework.

As shown in Fig. 2b, the modulation behavior changes qualitatively once four-phonon scattering is included. In contrast to the weak bidirectional response observed in the 3ph-only framework in Fig. 2a, the 3ph+4ph framework produces a predominantly suppressive modulation over the scanned frequency range. No robust enhancement window is retained after four-phonon scattering is taken into account.

For an excitation intensity of 5, the relative thermal conductivity remains nearly unchanged below 2 THz and within 9.6–17.6 THz, whereas clear suppression appears in the low-frequency region of 2.0–9.6 THz and the high-frequency region of 17.6–21.6 THz. The strongest suppression at this excitation intensity occurs at 20.5 THz, where $\kappa/\kappa_0$ decreases to 0.828. In addition, a broad low-frequency suppression band is observed around 4–6 THz, with representative local minima near 4.4 and 4.8 THz.

When the excitation intensity is increased to 25, the suppressive modulation is markedly amplified while the overall frequency dependence remains similar. The strongest suppression again appears at 20.5 THz, where $\kappa/\kappa_0$ drops to 0.415. Pronounced suppression is also observed throughout a wide low-frequency interval of approximately 2.2–9.2 THz and a high-frequency interval of approximately 18.0–21.3 THz. In particular, the low-frequency region around 4.1–5.8 THz and the high-frequency region around 18.4–20.5 THz show especially strong responses, indicating that these phonon windows are highly effective for suppressing heat transport in bulk BAs.

The comparison between Fig. 2a and Fig. 2b indicates that higher-order phonon scattering does not merely rescale the magnitude of thermal-conductivity modulation. Instead, represented here by four-phonon processes, it qualitatively changes the net response from weak bidirectional tuning in the 3ph-only framework to predominantly suppressive thermal-conductivity modulation in the 3ph+4ph framework. This contrast directly demonstrates the governing role of higher-order phonon scattering in targeted thermal modulation of bulk BAs and motivates the following scattering-rate analysis.

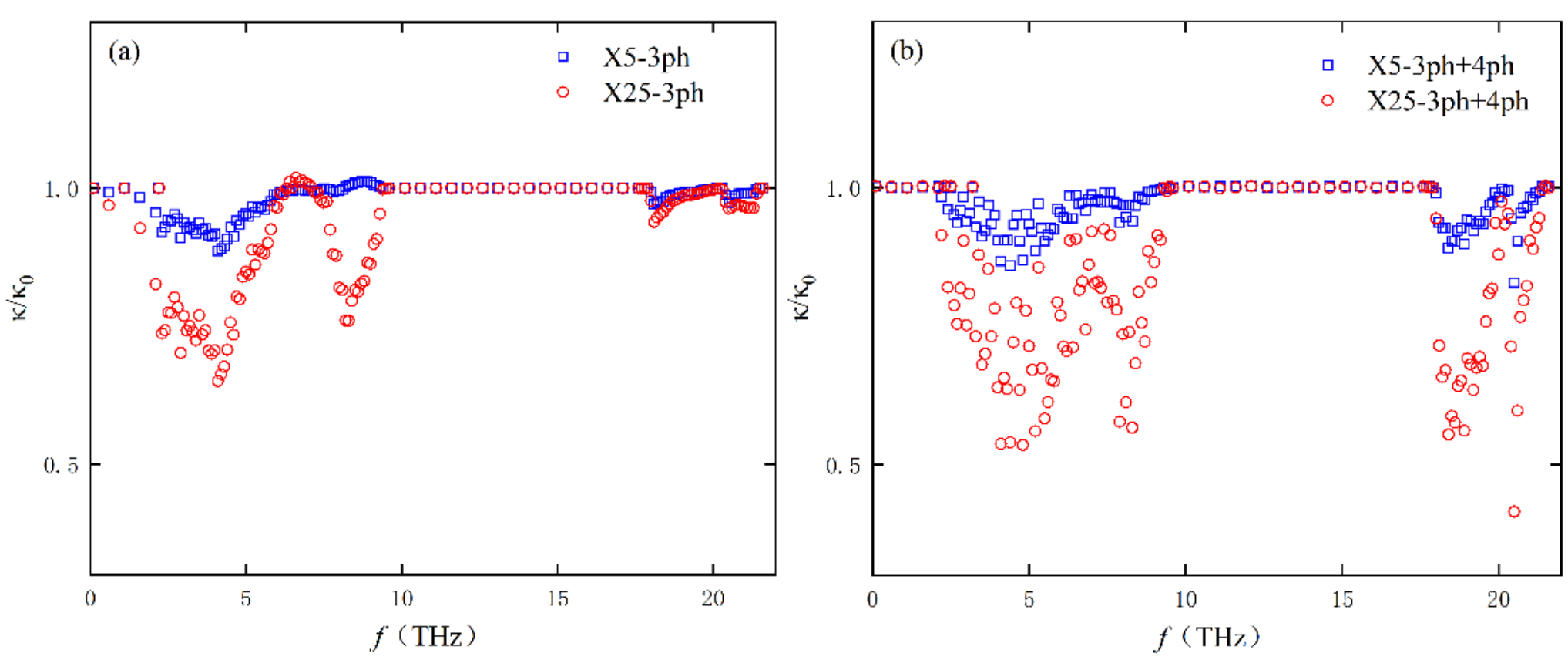

**Figure 2** Relative thermal conductivity $\kappa/\kappa_0$ of bulk BAs as a function of the center frequency of the targeted excitation window at 300 K for excitation intensities of 5 and 25. (a) Results obtained in the three-phonon-only framework. (b) Results obtained in the combined three-phonon plus four-phonon framework. In each calculation, only phonons within a 0.1 THz-wide frequency window were excited, while all other phonon modes retained their intrinsic distribution. Coarse scans with a 0.5 THz step were used over the full frequency range, while finer scans with a 0.1 THz step were performed in the strongly modulated regions. κ0 denotes the intrinsic thermal conductivity calculated in the corresponding framework, namely 2302 W/(m • K) for the 3ph-only framework and 1220 W/(m • K) for the 3ph+4ph framework.

To clarify why bulk BAs exhibits weak bidirectional modulation in the 3ph-only framework but predominantly suppressive modulation in the 3ph+4ph framework, we further compared their phonon-scattering characteristics. The distinct modulation behaviors do not originate from a simple rescaling of thermal conductivity, but rather from the combined effects of different intrinsic scattering backgrounds and different excitation-induced responses of the heat-carrying phonons.

The intrinsic phonon scattering-rate spectra at 300 K are compared in Fig. S5. After four-phonon scattering is included, the overall scattering rates are raised over a broad frequency range, and this increase is especially relevant in the low-frequency acoustic region that dominates heat transport. Therefore, even before targeted excitation is introduced, the 3ph+4ph framework already places bulk BAs in a stronger-scattering background than the 3ph-only framework. Under such a condition, the transport-promoting effect associated with the excitation-induced increase in phonon population becomes less competitive, whereas the scattering-enhancement effect becomes more dominant.

On this basis, we selected two representative excitation windows to examine how the scattering rates respond to targeted excitation in the two frameworks. The first case is the 8.2–8.3 THz window at an excitation intensity of 5, where the 3ph-only calculation gives a weak enhancement or nearly unchanged response, whereas the 3ph+4ph result becomes suppressive. As shown in Fig. 3a and Fig. 3b, the excitation-induced change in the scattering rates of low-frequency heat-carrying phonons remains relatively limited in the 3ph-only framework. As a result, the population increase of the excited phonons can still survive as a weak transport-promoting contribution. In contrast, in the 3ph+4ph framework, the same excitation acts on top of an already elevated intrinsic scattering background and produces a more systematic increase in the scattering rates of low-frequency heat-carrying phonons. Consequently, the scattering-enhancement effect outweighs the transport-promoting effect, and the net modulation changes from weak enhancement to suppression.

The second representative case is the 4.1–4.2 THz excitation window at an excitation intensity of 25, where both frameworks yield suppressive modulation. As shown in Fig. 3c and Fig. 3d, targeted excitation increases the scattering of low-frequency heat-carrying phonons in both the 3ph-only and 3ph+4ph frameworks, which explains why thermal conductivity is reduced in both cases. However, the excited-state scattering rates in the 3ph+4ph framework remain systematically higher over the low-frequency heat-carrying region. Therefore, although the modulation sign is suppressive in both frameworks, the suppression becomes much stronger once four-phonon scattering is included.

These results, together with the intrinsic scattering-rate comparison in Fig. S5, indicate that higher-order phonon scattering governs targeted control of heat conduction in bulk BAs through a dual role. On the one hand, four-phonon scattering raises the intrinsic scattering background of the system; on the other hand, it makes low-frequency heat-carrying phonons more susceptible to excitation-induced scattering enhancement. As a result, the weak enhancement window retained in the 3ph-only framework is removed once four-phonon scattering is included. Even in frequency windows where both frameworks remain suppressive, the 3ph+4ph framework still yields a substantially stronger reduction in thermal conductivity. Therefore, the predominantly suppressive modulation behavior of bulk BAs at room temperature originates from the governing role of higher-order four-phonon scattering in strengthening the scattering of low-frequency heat-carrying phonons.

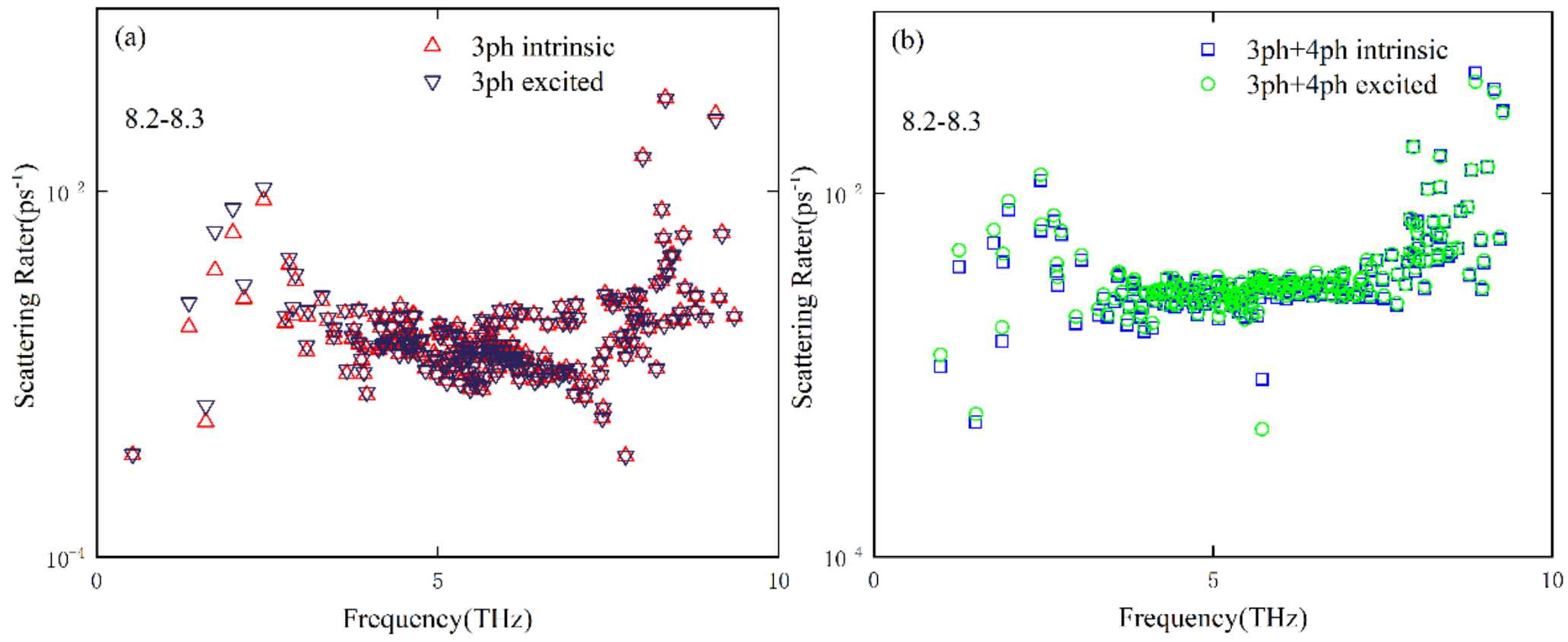

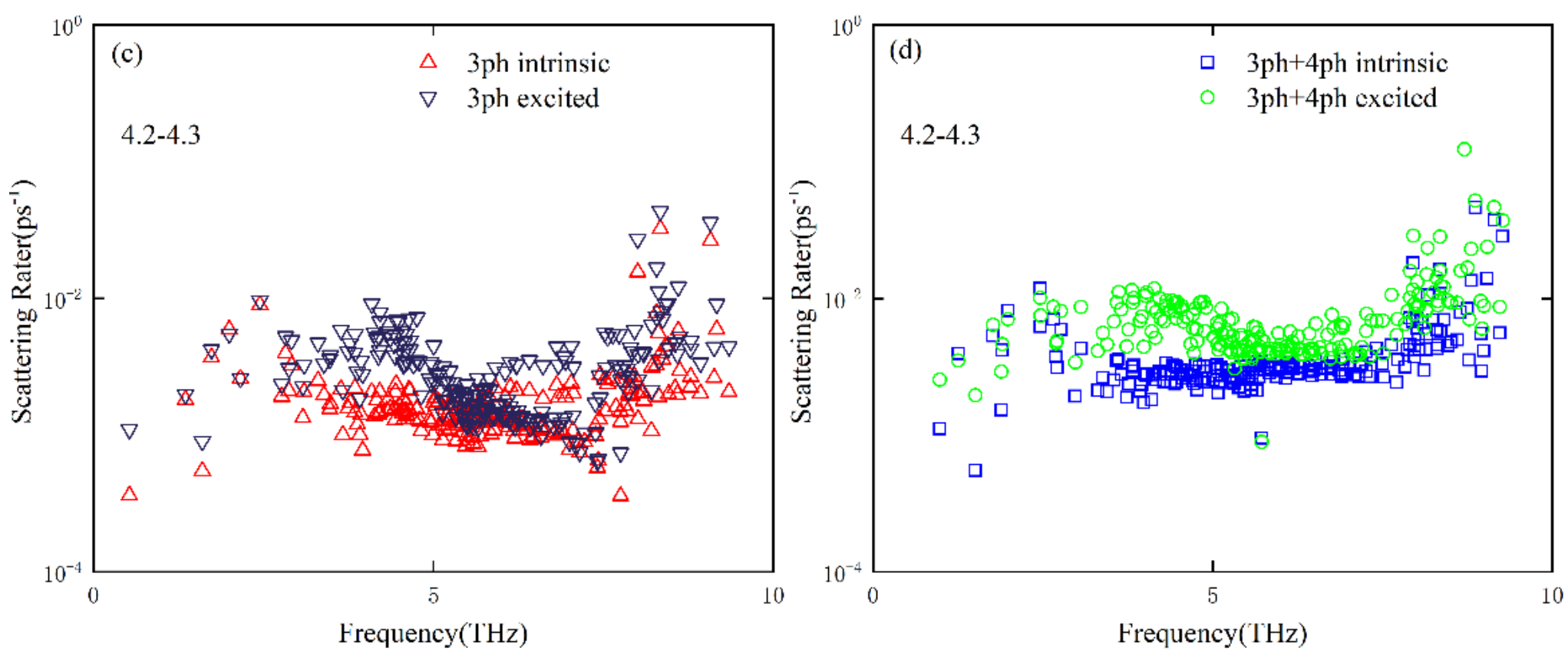

**Figure 3** Phonon scattering-rate spectra of bulk BAs at 300 K in the 0–10 THz range for the intrinsic and excited states in the 3ph-only and 3ph+4ph frameworks. (a) 8.2–8.3 THz frequency range at an excitation intensity of 5 for the 3ph-only framework. (b) 8.2–8.3 THz frequency range at an excitation intensity of 5 for the 3ph+4ph framework. (c) 4.1–4.2 THz frequency range at an excitation intensity of 25 for the 3ph-only framework. (d) 4.1–4.2 THz frequency range at an excitation intensity of 25 for the 3ph+4ph framework.

To further examine the temperature dependence of targeted-excitation modulation and the role of four-phonon scattering, we compared the 3ph+4ph results at 300 K and 100 K. The relative thermal conductivity under excitation intensities of 5 and 25 is shown separately in Figs. 4a and 4b, respectively. In addition, the ratio between the intrinsic four-phonon and three-phonon scattering rates, $\Gamma_{4ph}/\Gamma_{3ph}$, was analyzed in Fig. 4c to quantify the relative importance of four-phonon scattering at different temperatures.

As shown in Fig. 4a, at an excitation intensity of 5, lowering the temperature generally weakens the modulation effect in the 3ph+4ph framework. The 100 K curve remains closer to unity than the 300 K curve over most of the scanned frequency range. In the low-frequency region, the strongest suppression at 100 K occurs around 4.1 THz, where $\kappa/\kappa_0$ decreases to 0.905, whereas the corresponding value at 300 K is 0.864. Meanwhile, near the upper edge of the acoustic branches, weak enhancement reappears at 100 K. Around 8.1 THz, $\kappa/\kappa_0$ slightly exceeds unity at an excitation intensity of 5, while the corresponding 300 K result remains suppressive. These features indicate that the predominantly suppressive modulation observed at 300 K is weakened at lower temperature, and weak 3ph-like bidirectional characteristics can partially reappear.

When the excitation intensity is increased to 25, as shown in Fig. 4b, the suppressive response is amplified at both temperatures. However, the 300 K result still shows a much stronger suppression than the 100 K result. In the low-frequency region, excitation around 4.1 THz gives $\kappa/\kappa_0$ = 0.643 at 100 K and 0.540 at 300 K. The temperature dependence is even more evident in the high-frequency region. At 300 K, targeted excitation above 18 THz produces pronounced suppression, whereas at 100 K the relative thermal conductivity in the same frequency range remains much closer to unity. For example, under an excitation intensity of 25, excitation around 18.6 THz gives $\kappa/\kappa_0$= 0.578 at 300 K, but only weakly reduces it to 0.971 at 100 K. These results show that lowering the temperature does not simply reduce the modulation amplitude uniformly; instead, it strongly weakens the high-frequency suppressive response and partially restores the weak bidirectional features observed in the 3ph-only framework.

The microscopic origin of this temperature dependence is further clarified by the scattering-rate ratio shown in Fig. 4c. Compared with 100 K, the ratio $\Gamma_{4ph}/\Gamma_{3ph}$ at 300 K is significantly larger over the relevant frequency range, indicating that four-phonon scattering contributes much more strongly to the total anharmonic scattering at room temperature. This enhanced four-phonon contribution is especially important for the low-frequency acoustic phonons that dominate heat transport and for the high-frequency excitation windows that can affect low-frequency heat-carrying modes through phonon-phonon coupling. Therefore, at 300 K, targeted excitation acts on a stronger four-phonon scattering background, making the excitation-induced scattering enhancement more dominant than the transport-promoting effect associated with phonon population increase. Additional low-temperature calculations down to 10 K are provided in Fig. S6, showing that the weak enhancement near the 8.1–8.2 THz window appears only within an intermediate-temperature range and becomes nearly negligible at 10 K.

In contrast, at 100 K, the relative contribution of four-phonon scattering is substantially reduced, as reflected by the smaller $\Gamma_{4ph}/\Gamma_{3ph}$ ratio. Under this weaker four-phonon scattering background, the excitation-induced increase in scattering becomes less overwhelming. As a result, the high-frequency suppressive response is strongly weakened, and weak 3ph-like bidirectional features can reappear near the upper edge of the acoustic branches. These temperature-dependent results further support the mechanism discussed above: higher-order four-phonon scattering governs the targeted thermal-conductivity response of bulk BAs at room temperature, whereas lowering temperature weakens this higher-order scattering contribution and drives the modulation behavior closer to the 3ph-only picture.

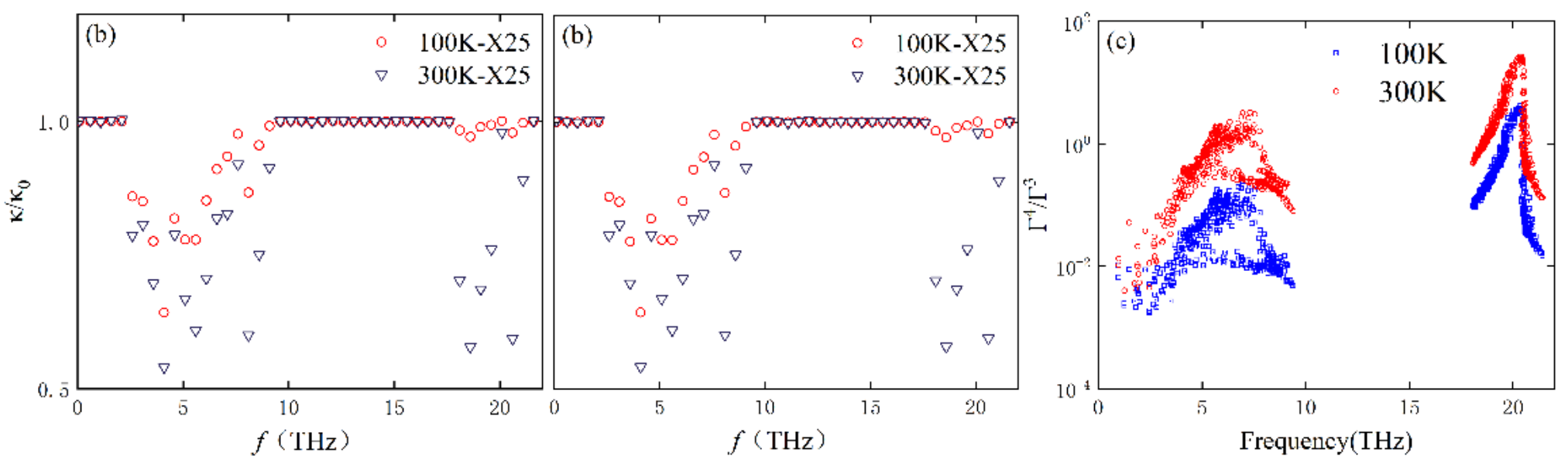

**Figure 4** Temperature-dependent modulation of thermal conductivity and relative four-phonon scattering strength in bulk BAs. (a, b) Relative thermal conductivity $\kappa/\kappa_0$ as a function of the center frequency of the targeted excitation window in the 3ph+4ph framework at 300 K and 100 K for excitation intensities of (a) 5 and (b) 25. In each calculation, only phonons within a 0.1 THz-wide frequency window were excited, while all other phonon modes retained their intrinsic distribution. (c) Intrinsic scattering-rate ratio $\Gamma_{4ph}/\Gamma_{3ph}$ as a function of phonon frequency at 300 K and 100 K. Only phonon modes with nonzero three-phonon scattering rates are included in the ratio plot.

## Conclusion

In this work, we demonstrated that higher-order phonon scattering governs targeted control of heat conduction in bulk boron arsenide (BAs). Based on first-principles calculations and phonon Boltzmann transport analysis, we found that targeted phonon excitation modulates the thermal conductivity of bulk BAs in a strongly frequency-dependent manner. Within the three-phonon-only framework, targeted excitation at 300 K produces a weak but clearly bidirectional modulation of thermal conductivity. However, once higher-order four-phonon scattering is included, the modulation changes qualitatively to a predominantly suppressive behavior. In the combined three-phonon plus four-phonon framework, the strongest suppression occurs at 20.5 THz, where $\kappa/\kappa_0$ decreases to 0.828 and 0.415 for excitation intensities of 5 and 25, respectively.

Mechanistically, higher-order four-phonon scattering plays a dual role in targeted thermal-conductivity modulation of bulk BAs. It raises the intrinsic scattering background and promotes a more systematic excitation-induced increase in the scattering of low-frequency heat-carrying phonons. As a result, the weak enhancement windows retained in the three-phonon-only framework are removed, and the net modulation becomes predominantly suppressive. Moreover, the comparison between 300 K and 100 K shows that lowering temperature weakens the predominance of four-phonon scattering and partially restores 3ph-like bidirectional features. These results demonstrate the feasibility of targeted phonon excitation in bulk materials and identify higher-order phonon scattering as a key factor governing the net modulation effect. This work provides useful guidance for phonon engineering and dynamic thermal management in three-dimensional systems.

## Acknowledgement

The authors are grateful to Xiao Wan for helpful discussions. The authors thank the National Supercomputing Center in Tianjin (NSCC-TJ) and the TianHe High Performance Computer for providing computational resources.

## Data Availability

The data that support the findings of this study are available from the corresponding author upon reasonable request.